\documentclass[aps,prb,twocolumn,letterpaper,superscriptaddress,floatfix,showpacs]{revtex4-1}

\usepackage{graphicx,amsmath,amsfonts,amssymb,color,ulem}

\def\PuCoIn{PuCoIn$_5$}

\def\PuCoGa{PuCoGa$_5$}
\def\PuRhGa{PuRhGa$_5$}
\def\PuRhIn{PuRhIn$_5$}

\def\iTone{$T_1^{-1}$}
\def\Tc{$T_c$}

\def\nuq{$\nu_Q$}
\def\ita{$\eta$}
\def\iTT{$\left(T_1T\right)^{-1}$}

\begin{document}

\title{Extended Nuclear Quadrupole Resonance Study of the Heavy-Fermion Superconductor \PuCoGa}

\author{G.~Koutroulakis}
\address{Los Alamos National Laboratory, Los Alamos, NM 87545}
\affiliation{Department of Physics $\&$ Astronomy, UCLA, Los Angeles, CA 90095, USA}
\author{H.~Yasuoka}
\address{Los Alamos National Laboratory, Los Alamos, NM 87545}
\author{P.~H.~Tobash}
\address{Los Alamos National Laboratory, Los Alamos, NM 87545}
\author{J.~N.~Mitchell}
\address{Los Alamos National Laboratory, Los Alamos, NM 87545}
\author{E.~D.~Bauer}
\address{Los Alamos National Laboratory, Los Alamos, NM 87545}
\author{J.~D.~Thompson}
\address{Los Alamos National Laboratory, Los Alamos, NM 87545}

\begin{abstract}

\PuCoGa\ has emerged as a prototypical heavy-fermion superconductor, with its transition temperature ($T_c\simeq18.5$ K) being the highest amongst such materials. Nonetheless, a clear description as to what drives the superconducting pairing is still lacking, rendered complicated by the notoriously intricate nature of plutonium's 5$f$ valence electrons. Here, we present a detailed $^{69,71}$Ga nuclear quadrupole resonance (NQR) study of \PuCoGa, concentrating on the system's normal state properties near to \Tc\ and aiming to detect distinct signatures of possible pairing mechanisms. In particular, the quadrupole frequency and spin-lattice relaxation rate were measured for the two crystallographically inequivalent Ga sites and for both Ga isotopes, in the temperature range \mbox{1.6 K -- 300 K}. No evidence of significant charge fluctuations is found from the NQR observables. On the contrary, the low-energy dynamics is dominated by anisotropic spin fluctuations with strong, nearly critical, in-plane character, which are effectively identical to the case of the sister compound \PuCoIn . These findings are discussed within the context of different theoretical proposals for the unconventional pairing mechanism in heavy-fermion superconductors.

\end{abstract}

\pacs{74.70.Xa,76.60.-k,74.25.nj,75.30.Kz}

\maketitle
\section{Introduction}

The character of the superconducting (SC) pairing in heavy-fermion (HF) compounds has remained a central open question, fitting in the broader puzzle of what mechanism drives unconventional superconductivity in general. The prevalent picture suggests that, in most cases (e.g. various cuprates \cite{Takigawa:1991,Chubukov2003}, iron-pnictides \cite{,Chubukov:2008,Ning:2009}, organics \cite{deSoto:1995,Ishiguro}, and HFs \cite{Mathur:1998}), spin fluctuations (SFs) associated with the proximity to an antiferromagnetic quantum critical point (QCP) provide the glue for the SC condensate. However, whether this magnetic mechanism is ubiquitous among unconventional SCs or some other mechanism could be playing an important role as well - like valence fluctuations (VFs) \cite{Miyake:2002jd,Miyake:2007gn} or even \textit{composite} pairing \cite{Flint:2008fk,Flint:2010hc}- is yet to be resolved.

One family of materials where this question is relevant is that of the heavy-fermion Pu-115 SCs, Pu$MX_5$ ($M$=Co,Rh, and $X$=Ga,In). The electronic properties of these materials are chiefly governed by the Pu 5$f$ electrons \cite{Booth2014}, which display a complex duality between itinerant and localized atomic-like behavior, leading to a variety of exotic highly correlated states, not the least of which is unconventional superconductivity. \PuCoGa\ becomes SC below \mbox{\Tc $\simeq$18.5 K} \cite{Curro:2005fw}, the highest critical temperature amongst heavy-fermion SCs and nearly an order of magnitude higher than the other Pu-115 family members: \PuRhGa\ (\mbox{\Tc $\simeq$8.7 K}) \cite{Wastin:2003wh},  \PuCoIn\ (\mbox{\Tc $\simeq$2.3 K}) \cite{Bauer:2011jl}, and \PuRhIn\ (\mbox{\Tc $\simeq$1.6 K}) \cite{Bauer:2015}.

There have been several attempts to draw a connection between the variation of $T_c$ and other physical quantities, such as a linear correlation to the lattice tetragonality $c/a$ \cite{Bauer:2004prl} or the spin-fluctuation energy anisotropy \cite{Baek:2010ca}, in an effort to delineate a common SC mechanism. Alternatively, one could hypothesize that the considerably higher \Tc\ of \PuCoGa\ is not simply a manifestation of a larger SC pairing energy scale, but rather the consequence of an entirely different pairing mechanism. One such plausible scenario for the Pu-115s would accommodate two distinctive SC domes in a potential $T-P$ phase diagram \cite{Bauer:2011jl}: One for the larger unit-cell volume (i.e., smaller effective chemical pressure) In compounds where SC is magnetically-mediated, near to an antiferromagnetic QCP; and another one, at higher effective chemical pressure, where \PuCoGa\ resides and where VFs help stabilize superconductivity proximate to a valence transition, similar to the case of CeCu$_2$Si$_2$ under pressure \cite{Yuan2104,Miyake:2002jd}. Further support for the latter scenario, at least intuitively, is the absence of a local magnetic moment in the normal state of \PuCoGa, i.e., it exhibits an approximately temperature-independent Pauli susceptibility \cite{Hiess:2008}, as well as the fact that it does not appear to be near a magnetically ordered state \cite{Boulet:2005}. 

Importantly, recent ultrasound spectroscopy measurements in \PuCoGa\ revealed an anomalous softening of the bulk modulus over a wide temperature range in the normal state, which is truncated upon entering the SC state \cite{Ramshaw2015}. The effect was attributed to strong fluctuations of the Pu 5$f$ mixed-valence state, which in turn drive the SC pairing thus avoiding a valence transition. Nevertheless, nuclear quadrupole and magnetic resonance (NQR and NMR, respectively) experiments have long provided evidence for the presence of strong antiferromagnetic SFs in the normal state of \PuCoGa\ and approaching \Tc\ \cite{Curro:2005fw,Baek:2010ca}, a hallmark of magnetically-mediated superconductivity. Hence, albeit challenging, probing directly for signatures of valence fluctuations and their relationship to any SFs could inform the conundrum of the SC pairing's detailed character in the Pu 115s. 

Here, we report an extended, comprehensive NQR study in \PuCoGa\ of both crystallographic Ga sites of the naturally abundant $^{69}$Ga and $^{71}$Ga, focusing on the temperature dependence of the electric field gradient (EFG) and the nuclear spin-lattice relaxation rate (\iTone) in the normal state. The ultimate goal is to provide insight on the presence of VFs, or lack thereof, and their possible relationship to the nature of normal-state SFs.

\section{Sample and Experimental details}

The sample consisted of approximately 100mg of powdered \PuCoGa\ crystals, synthesized as described in Ref.\cite{Bauer:2012_growth}. In order to prevent any radioactive contamination, the cylindrical NMR coil, with dimensions of 3 mm diameter and 7 mm length, was encapsulated in a Stycast 1266 epoxy $20 \rm{mm}\times20 \rm{mm}\times20 \rm{mm}$ mold, prior to sample insertion. The mold was drilled along the coil's axis and, upon sample insertion, its ends were sealed by \mbox{2 $\mu$m} diameter-pore titanium frits, in order to ensure good thermal contact with the variable temperature insert's $^4$He gas flow.

The NQR spectra were recorded using a commercial pulsed NMR spectrometer, after standard Hahn spin-echo pulse sequences for the $^{69}$Ga and $^{71}$Ga nuclear spins. The nuclear spin-lattice relaxation rate, \iTone, was measured after inversion of the nuclear magnetization by an rf-pulse and inspecting the recovery profile. For the Ga nucleus ($I=3/2$), the relevant nuclear transition is $\left<\pm 1/2\leftrightarrow\pm 3/2\right>$, and the time evolution of the nuclear magnetization, $M(t)$, is given by $M(t)=M(0)\left(1-2e^{-3t/T_1}\right)$, where $M(0)$ is the thermal equilibrium value and $t$ is the delay time after the inversion pulse.

\section{Results and Analysis}
\subsection{Temperature dependence of \nuq}

For isotopes with nuclear spin $I\geq3/2$, the NQR spectrum is determined by the interaction between the nuclear quadrupole moment ($eQ$) and the electric field gradient (EFG) at the nuclear site due to the non-spherical charge distribution ($eq$) of the electronic environment. The pertinent NQR Hamiltonian is 
\begin{equation}
\mathcal{H}_Q=\frac{h \nu_Q}{6} \left[ 3 \hat{I}_z^2 -  \hat{I}^2 + \frac{1}{2}\eta(\hat{I}_{+}^2 + \hat{I}_{-}^2) \right].
\label{NQR}
\end{equation}
The spin operators $I_\alpha$ are defined along the EFG's principal axes, the quadrupole frequency $\nu_Q$ is \mbox{$\nu_Q \equiv 3e^2qQ/(h 2I (2 I -1) )$}, while the EFG tensor components are incorporated in $eq$ and \ita\ as $eq=V_{zz}$, and $\eta \equiv |V_{XX} -V_{YY}|/|V_{ZZ}|$. The EFG component $V_{ZZ}$ is taken by convention  to have the largest magnitude, and $V_{XX}$,$V_{YY}$ are chosen so that $0\leq\eta\leq1$.  In the case of $I=3/2$, Eq. \ref{NQR} results in a single NQR line at frequency $\nu_{\rm{NQR}}$, which can be expressed as
\begin{equation}
\nu_{NQR}=\nu_Q\sqrt{1+\frac{\eta^2}{2}}
\label{eq:NQR_3o2}
\end{equation}

\begin{figure}[htb]
\includegraphics[width=3.3in]{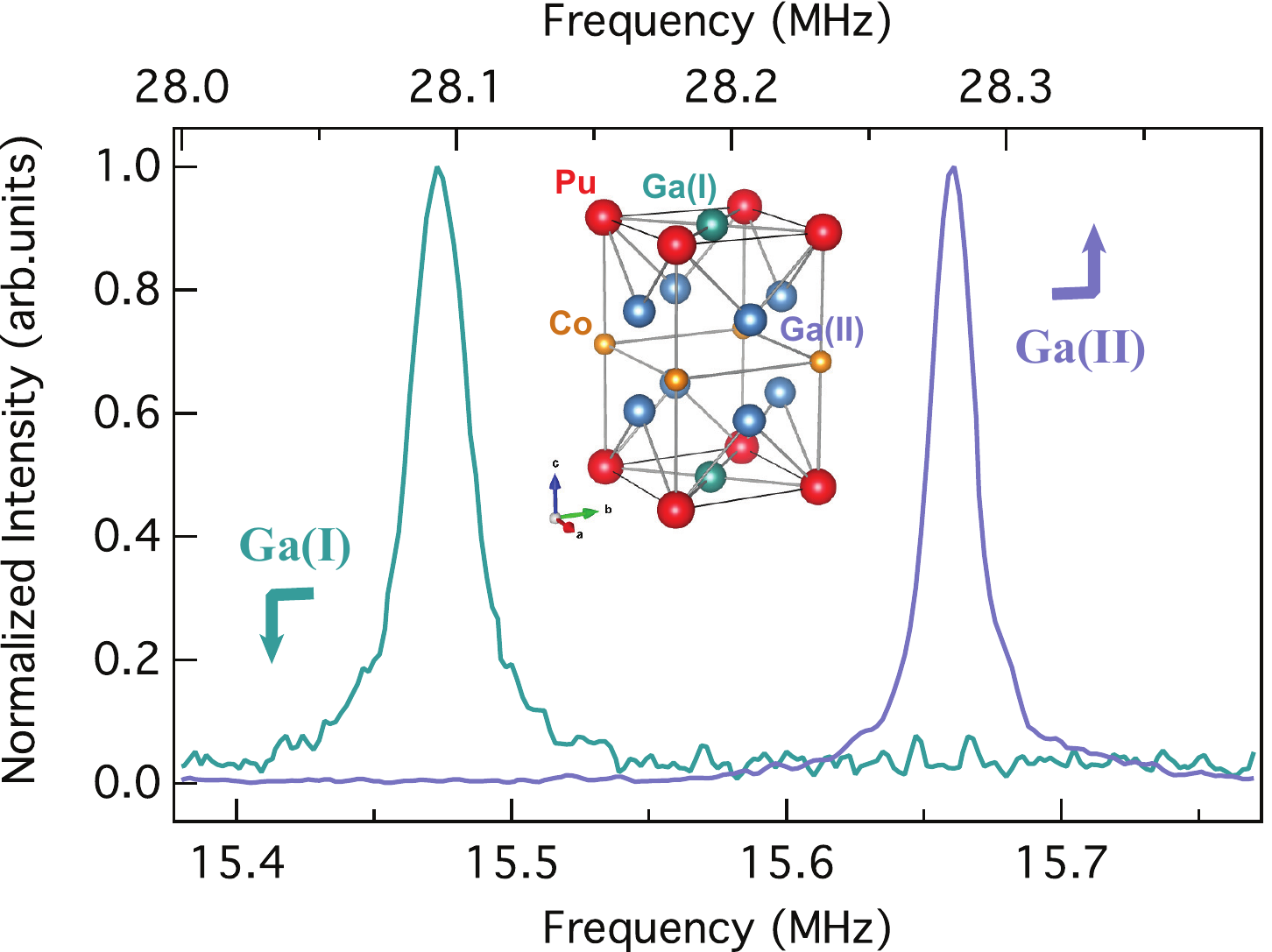}
\caption{NQR spectrum of the $^{69}$Ga(I) (lower axis) and $^{69}$Ga(II) (upper axis) sites, at $T=20$ K. \textit{Inset}: Unit cell structure of \PuCoGa . }
\label{fig:Spec}
\end{figure}

The crystal structure of \PuCoGa , shown in the inset of Fig. \ref{fig:Spec}, comprises alternating layers of PuGa$_3$ and CoGa$_2$ stacked along the $c$-axis, thus adopting the HoCoGa$_5$ tetragonal structure with space group $P4/mmm$. There are two crystallographically inequivalent Ga sites in the unit cell, henceforth labeled as Ga(I) and Ga(II). The Ga(I) site is situated in the middle of the $ab$-plane (1c site) and it has uniaxial symmetry ($V_{zz}\parallel\hat{c}$, $\eta=0$), whereas the Ga(II) site, sitting on the face of the unit cell (4i site), displays lower symmetry  ($V_{zz}\parallel\hat{a}\ \rm{or}\ \hat{b}$, $\eta\neq0$). Also, there are two NQR active Ga isotopes, $^{69}$Ga and $^{71}$Ga, with distinct quadrupole moment,  $eQ=0.178$ barn and $eQ=0.112$ barn respectively, which give rise to separate NQR lines.

The $^{69}$Ga NQR signal of the Ga(I) and Ga(II) sites was followed in the normal state of \PuCoGa\ and up to $T$=300 K. Representative spectra for both sites are depicted in Fig. \ref{fig:Spec}, for $T$=20 K. The assignment of each NQR peak to the respective Ga site is informed by the their relative intensity ratio, which should be 1:2 per site occupancy in the unit shell \cite{CurroNote}. Furthermore, it reproduces effectively the Ga NMR frequencies of the spectra featured in Ref. \cite{Baek:2010ca}, and it is in agreement with the values predicted by band structure calculations of the EFG which yield $^{69}\nu_{NQR}=19.5$ MHz, $\eta=0$ and $^{69}\nu_{NQR}=28.07$ MHz, $\eta=0.27$ for Ga(I) and Ga(II), respectively \cite{Harima}.

\begin{figure}[htb]
\includegraphics[width=3.3in]{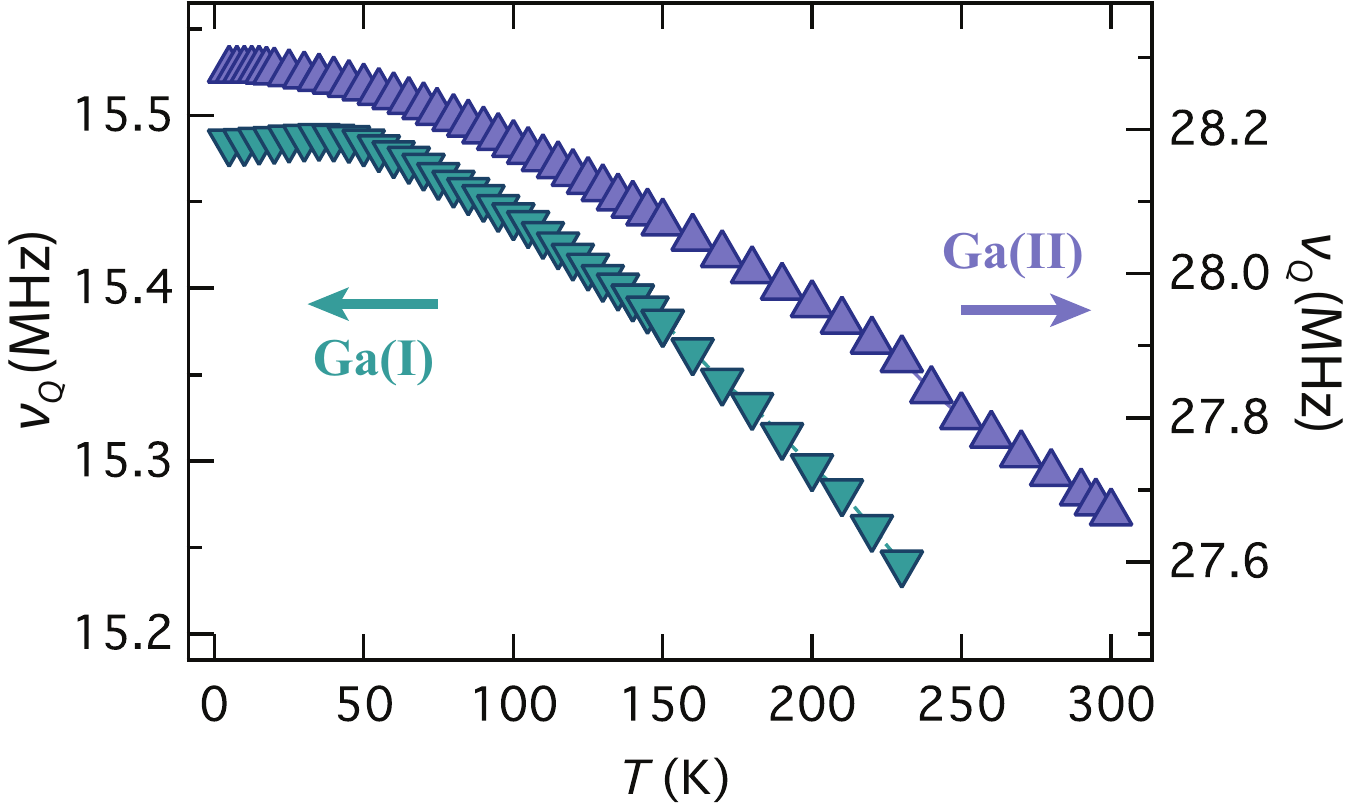}
\caption{Temperature dependence of the quadrupole frequency, $\nu_Q$, for $^{69}$Ga(I) (left axis, down triangles) and $^{69}$Ga(II) (right axis, up triangles), in the normal state. }
\label{fig:vq_vsT}
\end{figure}

The temperature evolution of the quadrupole frequency, $\nu_Q(T)$, is plotted in Fig. \ref{fig:vq_vsT} for both $^{69}$Ga sites, as extracted from the NQR peak position according to Eq. \ref{eq:NQR_3o2}. The spectral linewidth (not shown) remains  nearly unchanged throughout the probed temperature range. For increasing temperature, \nuq\ decreases, as generally expected due to the EFG being coupled to the lattice expansion. Deriving a relevant analytical expression for $\nu_Q(T)$ is not a straightforward task here, since the EFG originates with the onsite contribution of the electronic orbital wave function at the nuclear site. Instead, the following empirical formula, valid for conventional non-cubic metals \cite{Christiansen:1976wh}, is adopted:
\begin{equation}		 											 
\nu_Q(T)=\nu_Q(0)\left(1-A\cdot T^{3/2}\right),\ A>0\ .
\label{nuq}
\end{equation}

\begin{figure}[htb]
\includegraphics[width=3.3in]{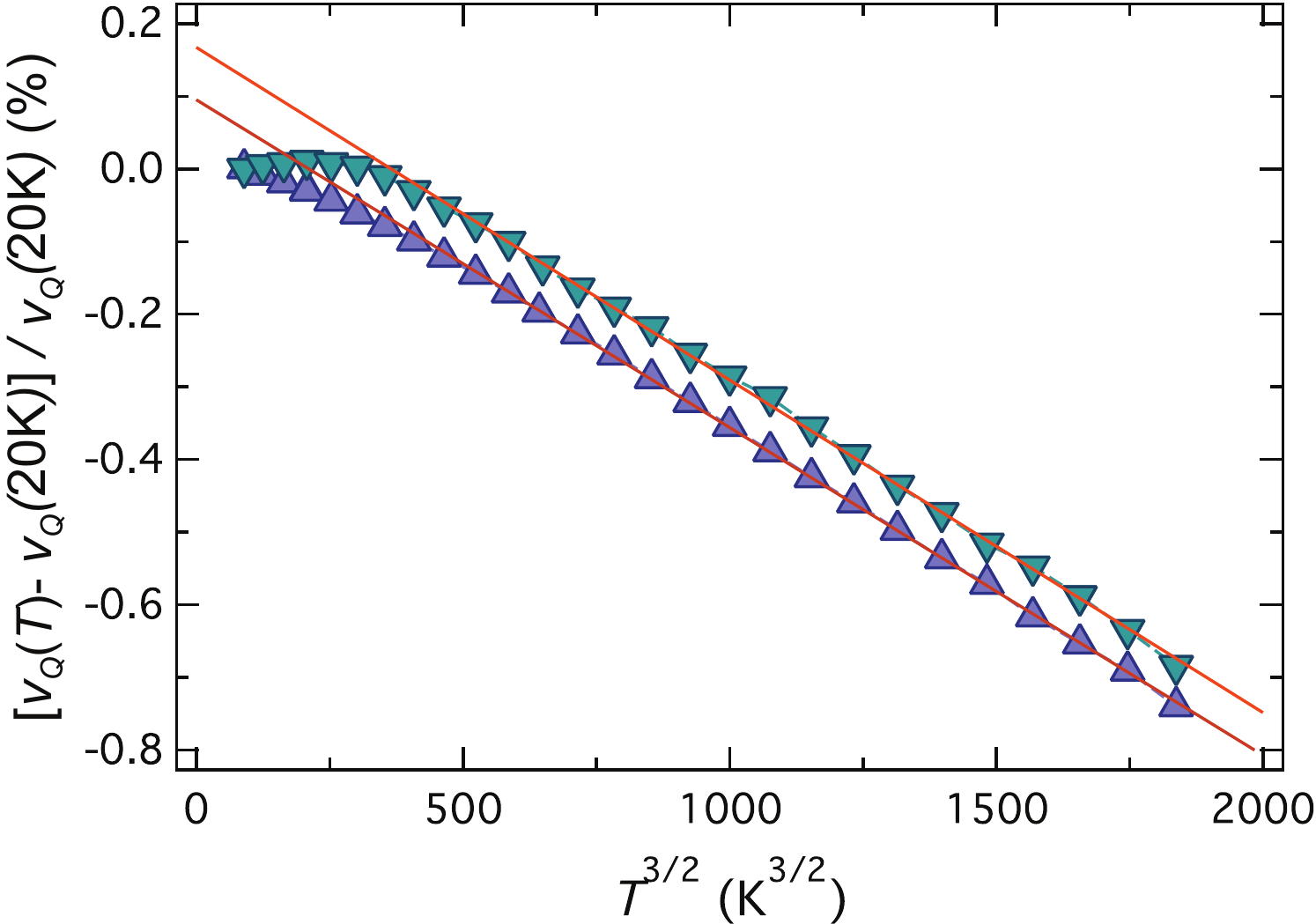}
\caption{Normalized fractional change of \nuq\ \textit{vs.} $T^{3/2}$ for $^{69}$Ga(I) (down triangles) and $^{69}$Ga(II) (up triangles), in the normal state. The solid lines illustrate linear fits to the data above $T=60$ K.}
\label{fig:vq_vsTsc}
\end{figure}

Fig. \ref{fig:vq_vsTsc} shows the fractional change of $\nu_Q(T)$ normalized at $T$=20K, just above \Tc, as a function of $T^{3/2}$. A fit to the data above 60K, per eq.\ref{nuq}, yields the values: $\nu_Q(0)$=15.519 MHz, $A=7.76\times10^{-5}$ K$^{-3/2}$ for Ga(I) and $\nu_Q(0)$=28.302MHz, $A=1.24\times10^{-4}$ K$^{-3/2}$ for Ga(II). While the empirical expression of Eq. \ref{nuq} describes well \nuq's temperature dependence above $T\simeq50$ K, a clear deviation sets in at lower temperature, most apparent for the in-plane Ga(I) site. This behavior is not fully understood at the moment, but it should be noted that it somewhat resembles the reported anomalous softening of the bulk modulus \cite{Ramshaw2015}. This raises the possibility that the observed \nuq\ temperature dependence is the result of an unusual EFG variation, through coupling to valence fluctuations of the Pu 5$f$ moments. What is more, the anisotropic character of these fluctuations, as reflected on the anisotropy between the in- and out-of-plane Poisson's ratio\cite{Ramshaw2015}, would agree with the in-plane Ga(I) site's \nuq\ being more readily affected than that for Ga(II), as observed.

\subsection{Temperature dependence of \iTone}

\begin{figure}[htb]
\includegraphics[width=3.3in]{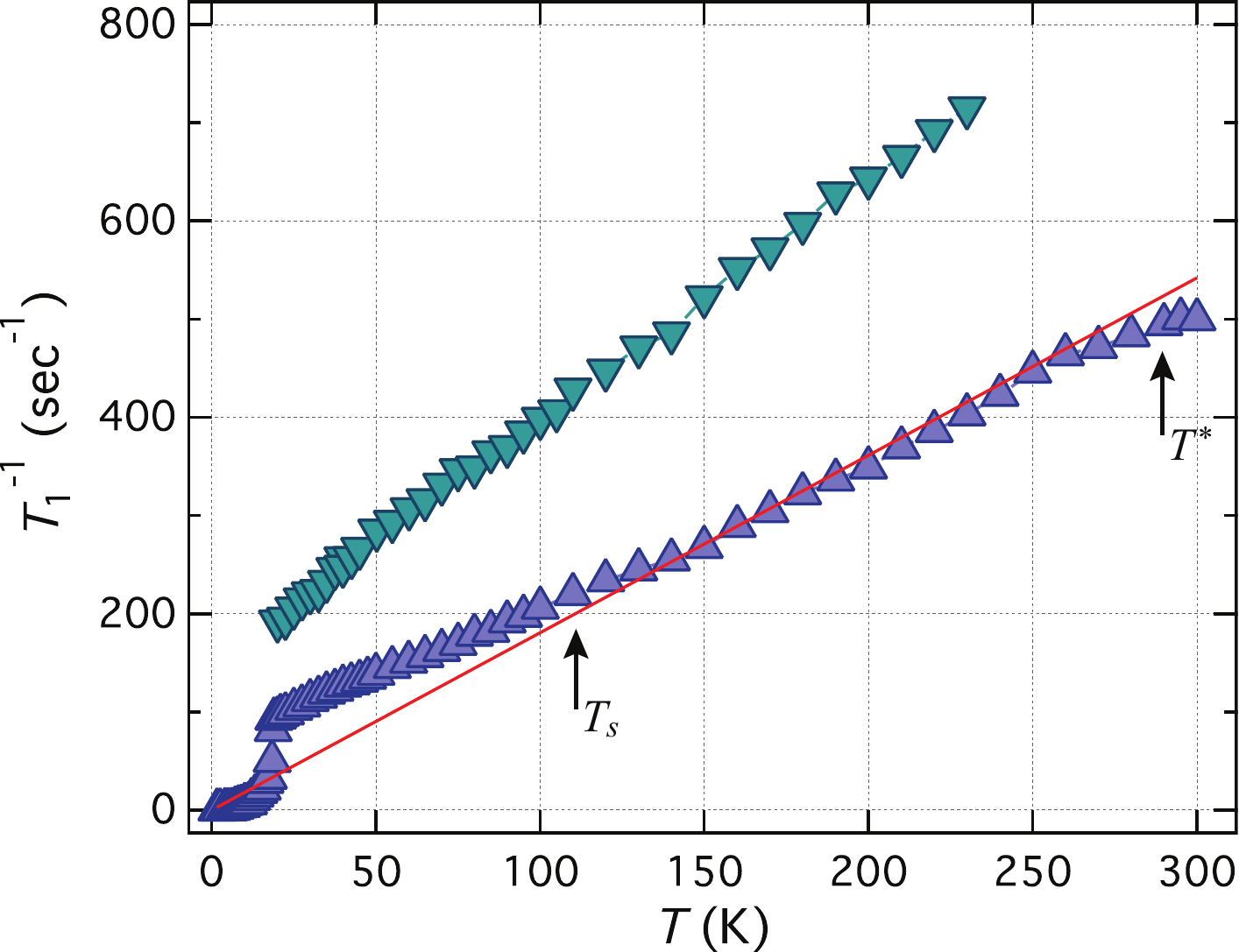}
\caption{Spin-lattice relaxation rate, \iTone , of $^{69}$Ga(I) (down triangles) and $^{69}$Ga(II) (up triangles) as a function of temperature. The solid line denotes the $T$-linear behavior, expected for a Fermi liquid.}
\label{fig:invT1_vsT}
\end{figure}

The relaxation rate \iTone\ was measured up to 230 K for Ga(I) and 300 K for Ga(II), with the results being plotted in Fig. \ref{fig:invT1_vsT}. In the normal state, there are three distinctive regimes, clearly observed in the Ga(II) data, characterized by the crossover temperatures $T^{\star}$ and $T_s$. At high $T$, a weakly interacting local-moment behavior gives way to the coherent heavy Fermi-liquid regime below $T^{\star}\sim285$ K, which corresponds to the Kondo coherence temperature in \PuCoGa\ as identified by resistivity measurements \cite{Fisk:1988fm,Bauer:2004prl}. Above $T^{\star}$, \iTone\ is temperature independent as expected for exchange-coupled local moment fluctuations, while it displays a typical Korringa-like behavior, $T_1^{-1}\sim T$,  below $T^{\star}$, where the relaxation process is governed by electron-hole pair excitations across the Fermi level. 

Going to lower temperature, below $T_s\sim110$ K, \iTone\ is enhanced beyond the $T$-linear expectation, hinting to the emergence of strong SFs, which have long been perceived as driving the SC pair formation \cite{Mathur:1998,Curro:2005fw,Kambe:2007jm}. A rapid decrease is seen upon entering the SC state below \mbox{$T_c\simeq 18.5$ K}, as the SC gap develops, and the temperature evolution of \iTone\ is that of an unconventional nodal-gap superconductor \cite{Curro:2005fw,Sakai:2005fh,Koutroulakis:2014}.

\begin{figure}[htb]
\includegraphics[width=3.5in]{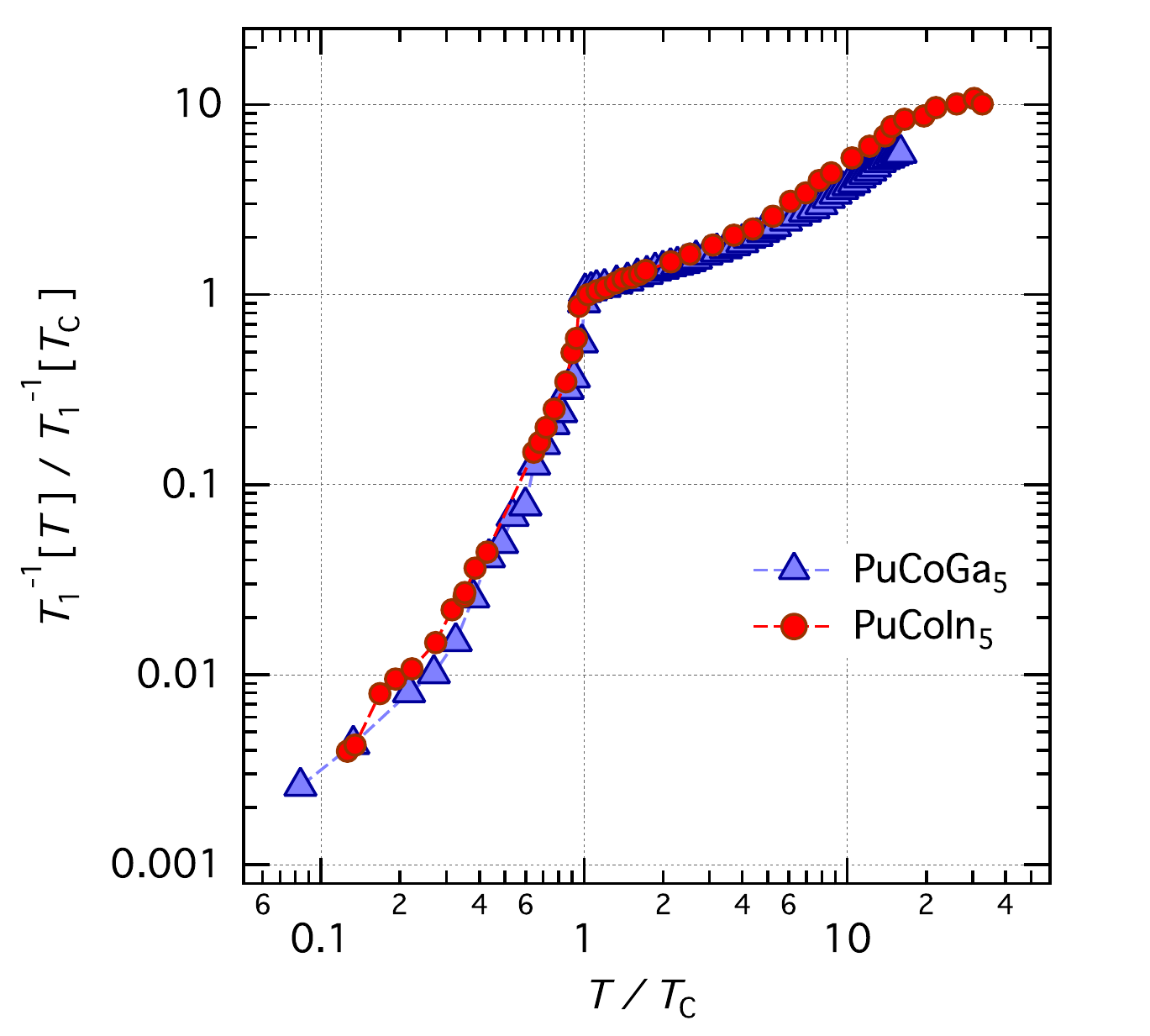}
\caption{Comparison of the relaxation rate in \PuCoGa\ and \PuCoIn\ (Ga(II) and In(II) sites), normalized by the different transition temperature \Tc . As discussed in the text, the temperature evolution of \iTone\ in the two materials is nearly identical, when scaled with \Tc\ (see also Table above) . Data for \PuCoIn\ are adopted from Ref. \cite{Koutroulakis:2014} .}
\label{fig:invT1_comp}
\end{figure}

Strikingly, the overall behavior of \iTone\ \textit{vs.} temperature in \PuCoGa, when scaled with \Tc,  is identical to that in its sister compound \PuCoIn, with the lower \mbox{\Tc $\simeq$2.3 K}, as demonstrated in Fig. \ref{fig:invT1_comp}. This indicates that the nature of SFs in the two materials is effectively the same, at least as probed by NQR relaxation, which would in turn favor as more likely a common underlying mechanism for superconductivity, despite the relatively big difference in \Tc.

\begin{table}[h]
\begin{tabular}{l|cccc}
\hline
compound \ &\  $V$ (\AA$^3$) \ &\  $T_c$ (K)\  &\  $T^{\star}$ (K)\  &\  $T_s$ (K)\   \\
\hline
\PuCoGa\ & 122  & 18.5 & 285 & 110  \\
\PuCoIn\ & 156 & 2.3 & 50  & 12  \\
\hline
\end{tabular}
\label{tab:CharTemp}
\caption{Characteristic temperatures in the PuCo$X_5$ compounds ($X$=Ga,In), as manifested in the NQR relaxation measurements (see text): $T_c$ is the SC critical temperature, $T^{\star}$ signals the Kondo coherence temperature, and $T_s$ corresponds to the onset of strong SFs nearing $T_c$. The unit cell volume $V$ is also listed.}
\end{table}

\begin{figure}[htb]
\includegraphics[width=3.2in]{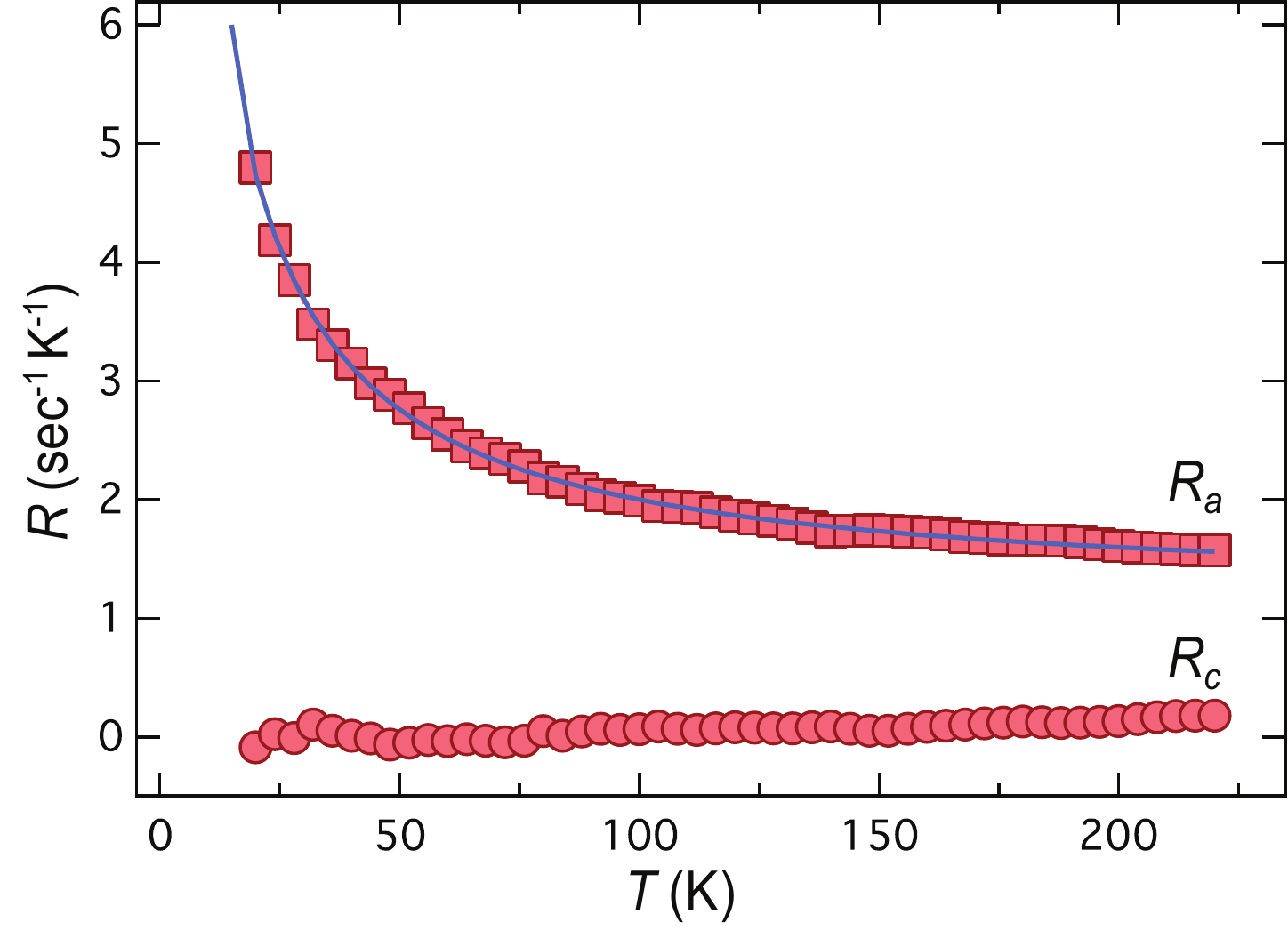}
\caption{Temperature dependence of the direction-specific relaxation rates $R_a$ and $R_c$. The solid line corresponds to a Curie-Weiss behavior fit for the in-plane rate, as discussed in the text.}
\label{fig:Rac}
\end{figure}

Generally, $T_1^{-1}(T)$ due to magnetic excitations can be expressed in terms of the imaginary part of the dynamic spin susceptibility $\chi(\mathbf{q},\omega)$ and the hyperfine coupling constant $A(\mathbf{q})$ as \cite{Moriya:1963hp},
\begin{equation}
\left(\frac{1}{T_1T}\right)_{\parallel}  \propto\sum_{\mathbf{q}}\left[\gamma_n A_{\perp}(\mathbf{q})\right]^2\frac{\chi_{\perp}''(\mathbf{q},\omega_0)}{\omega_0} ,
\label{eq:moriya}
\end{equation}
where $\gamma_n $ is the gyromagnetic ratio of the nucleus (for $^{69,71}$Ga,  $^{69}\gamma_n $=10.22 MHz/T  and $^{71}\gamma_n $=12.984 MHz/T),  $\omega_0$ is the Larmor frequency, and $\parallel$ ($\perp$)  corresponds to the direction parallel (perpendicular) to the quantization axis. Given the crystal structure of \PuCoGa, the nuclear spin quantization axis is $\hat{c}$ and $\hat{a}$ (or $\hat{b}$) for the Ga(I) and Ga(II) sites, respectively. Hence, the Ga(I) relaxation rate is sensitive only to the in-plane $\chi(\mathbf{q},\omega)_{a,b}$, while that of Ga(II) depends both on $\chi(\mathbf{q},\omega)_{a,b}$ and $\chi(\mathbf{q},\omega)_{c}$. We can then define direction-specific rates that probe solely the in- or out-of-plane component of the fluctuations in terms of the measured relaxation for the two Ga sites \cite{Kambe:2007jm,Koutroulakis:2014}. These are given by (assuming $A_{\rm{Ga(I)}}\sim A_{\rm{Ga(II)}}$)
\begin{equation}
 R_a=\frac{1}{2}\left(T_1T\right)^{-1}_{\rm{Ga(I)}},\  R_c=\left(T_1T\right)^{-1}_{\rm{Ga(II)}}-\frac{1}{2}\left(T_1T\right)^{-1}_{\rm{Ga(I)}}\ ,
\label{eq:Rac}
\end{equation}
and their evolution with temperature is shown in Fig. \ref{fig:Rac}. The out-of-plane component $R_c$ takes a very small, nearly temperature-independent value, while the considerably larger in-plane rate $R_a$ displays a rapid increase with lowering temperature and approaching \Tc. This suggests a strongly anisotropic character for the system's SFs, with their characteristic enhancement nearing the SC transition being dominated by the in-plane component, which, again, closely resembles the findings in \PuCoIn\ \cite{Koutroulakis:2014} as well as several other heavy-fermion SCs \cite{Kambe:2007jm,Baek:2010ca}. The same conclusion was previously reached by analysis of $^{59}$Co NMR relaxation data \cite{Baek:2010ca}.

\section{Discussion}
	
The central question remains whether the NQR results can provide any insight into the mechanism responsible for the unconventional superconductivity in \PuCoGa, and in the Pu$MX_5$ family more broadly. In general, the strong SFs  seen in numerous unconventional SCs close to \Tc, evidenced for example by the enhanced NQR relaxation rate, have posed as a likely candidate for mediating the Cooper-pair formation. The often observed critical character of these fluctuations, especially, in conjunction with the demonstrated proximity to magnetically ordered ground states, have made a strong case for magnetically mediated superconductivity. Interestingly, we find here, as shown in Fig. \ref{fig:Rac}, that the in-plane SFs diverge at low temperature in the normal state. In fact, fitting the in-plane $R_a(T)$ to a Curie-Weiss behavior plus a constant offset (solid blue line) yields $R_a(T)=1.18+\frac{85.4}{T+4.05}$. The small value of the Curie-Weiss temperature suggests critical behavior for the system's SFs, which indicates that \PuCoGa\ should indeed be near to an antiferromagnetic QCP. Nevertheless, attempts to reveal this putative neighboring magnetic state by chemical substitution have failed \cite{Boulet:2005}, casting doubt on its existence. Furthermore, ultrasound spectroscopy measurements have found an anomalously anisotropic behavior of the Poisson's ratios, attributed to 2D-like, in-plane strong valence fluctuations. With that in mind, one could hypothesize alternatively that the presumed QCP in \PuCoGa\ is associated with a valence transition, and the detected SFs by the NQR relaxation are the result of an intricate coupling between dynamic charge and spin susceptibility,  for example via spin-flip processes \cite{Miyake:2007gn}.

\begin{figure}[htb]
\includegraphics[width=3.5in]{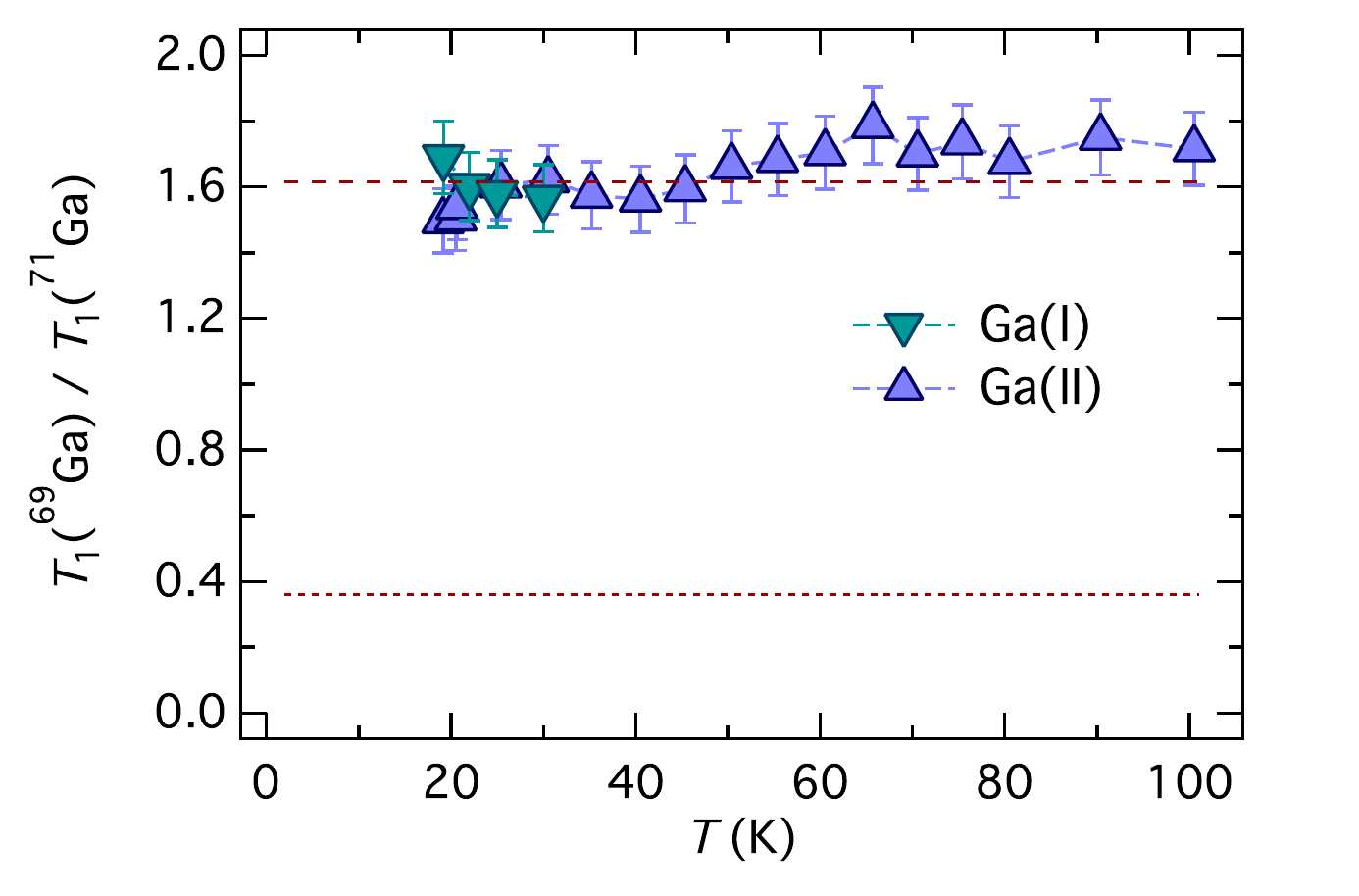}
\caption{Ratio of the relaxation rate of the Ga isotopes \textit{vs.} temperature, in the normal state. The horizontal dashed line at $\sim$1.6 corresponds to the ratio  $\left(^{71}\gamma_n/^{69}\gamma_n\right)^2$, while the one at $\sim$0.4 to $\left(^{71}Q/^{69}Q\right)^2$. }
\label{fig:T1ratio}
\end{figure}

To test this hypothesis, we looked specifically for signatures of normal-state charge (valence) fluctuations in the NQR relaxation data. First, in the case of critical valence fluctuations, it is theoretically predicted that \iTT\ should display a power-law variation $\left(T_1T\right)^{-1}\sim T^{-\zeta}$, with $0.5\leq \zeta\leq0.7$. \cite{Miyake:2014} Such a behavior has indeed been observed previously in several Yb-based compounds \cite{Ishida:2002,Deguchi:2012}, but it is absent in \PuCoGa. Second, if the NQR relaxation mechanism were dominated by charge fluctuations, this should be reflected on the ratio of the rates of the two Ga isotopes: The relaxation is generally governed solely by magnetic fluctuations, in which case the rates of different isotopes scale with the square of the respective gyromagnetic ratio (see Eq. \ref{eq:moriya}). For Ga nuclei, it is $\left(^{71}\gamma_n/^{69}\gamma_n\right)^2=1.614$. Nevertheless, if strong charge fluctuations are present and central to the relaxation process, the \iTone\ isotope ratio is modified and approaches that of the nuclear quadrupole moment $Q$ squared \cite{Slichter,Takagi:2004}, which in our case is $\left(^{71}Q/^{69}Q\right)^2=0.361$. The relaxation rate isotope ratio $T_1^{-1}(^{71}\rm{Ga})/T_1^{-1}(^{69}\rm{Ga})$ is plotted in Fig. \ref{fig:T1ratio} as a function of temperature for both Ga sites in \PuCoGa, in the normal state. No signature of prominent charge fluctuations is detected, since the system's low-energy dynamics appear to be fully dominated by magnetic fluctuations throughout the temperature range probed ($T_c\leq T\leq 100$ K).

All in all, our $^{69,71}$Ga NQR measurements do not show any evidence for the presence of charge fluctuations in the normal state of \PuCoGa . What is more, the detected enhanced relaxation confirms the previously observed, strongly in-plane SFs, which are effectively of identical character and magnitude to those in \PuCoIn\ (see Fig. \ref{fig:invT1_comp}),  suggesting a common SC pairing mechanism in these materials. In light of the ultrasound spectroscopy results \cite{Ramshaw2015}, we can not eliminate the possibility of charge fluctuations being present but not contributing significantly to the NQR observables, or being dominated by the effect of SFs. Given the qualitative similarity between the NQR and ultrasound findings, one could envision a scenario where spin and charge degrees of freedom are intricately coupled due to the strong hybridization, with the relevant strong 2D fluctuations being manifested differently in the two probes while being central to the formation of the SC condensate. Such a picture for the normal-state fluctuations in \PuCoGa\ would resemble, for example, the case of $\delta$-Pu where the mixed-valence 5$f$ states \cite{Booth2014} are accompanied by well-defined spin fluctuations, as was recently revealed by inelastic neutron scattering \cite{Janoschek:2015}.

\begin{figure}[ht]
\includegraphics[width=3.3in]{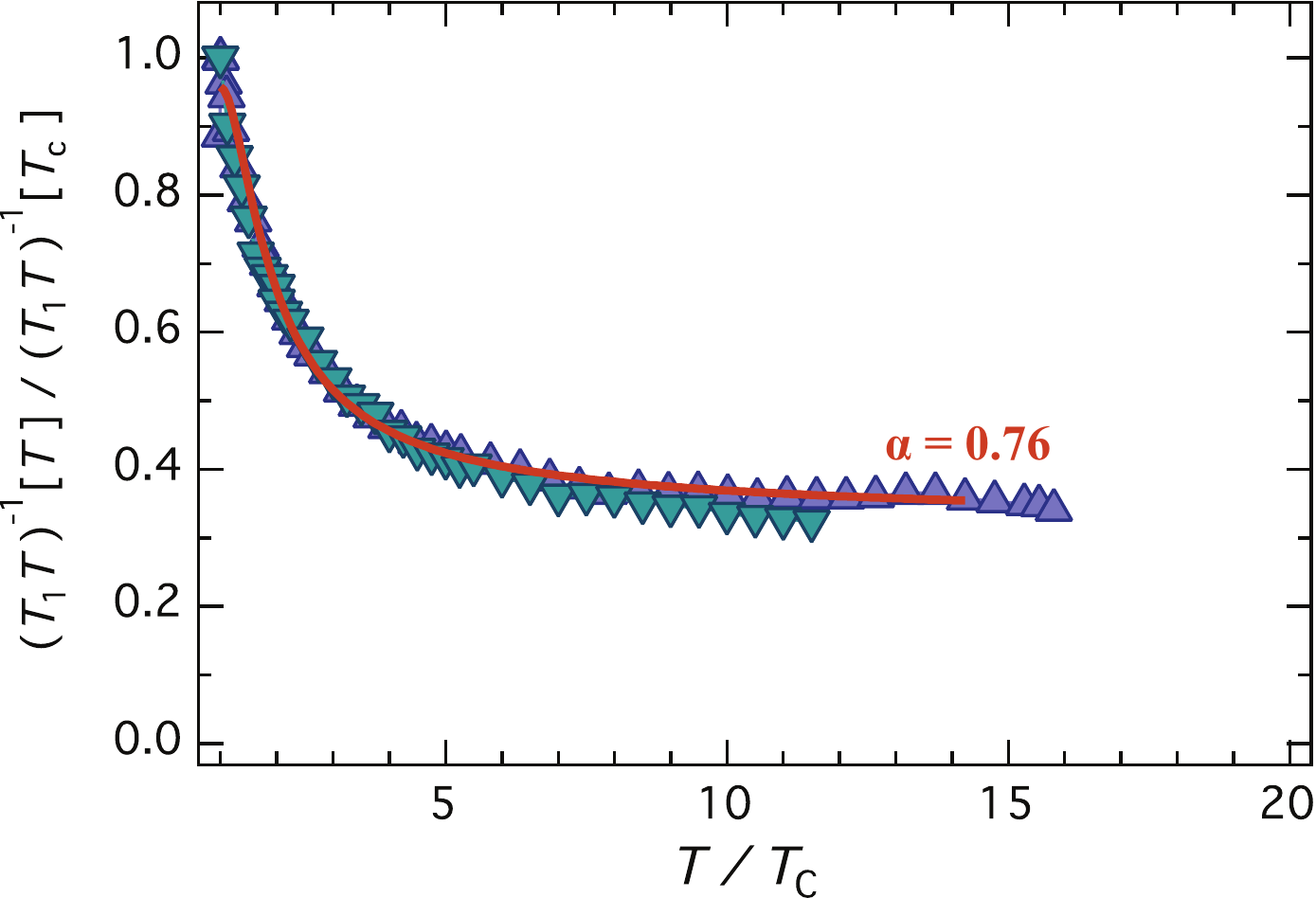}
\caption{Temperature dependence of \iTT , normalized to its value at \Tc , for $^{69}$Ga(I) (down triangles) and $^{69}$Ga(II) (up triangles), in the normal state. The solid curve describes the predicted form for the relaxation rate by the theory of composite pairing, as discussed in the text, for parameter $\alpha = 0.76$. }
\label{fig:invT1T}
\end{figure}

An alternative theoretical approach puts forth a \textit{composite} pairing mechanism, within the framework of  a two-channel Anderson model \cite{Flint:2008fk,Flint:2010hc,Flint:2011cl}, aiming to provide an overarching microscopic description for  heavy-fermion superconductivity. This theory predicts a sharp change in \nuq\ upon entering the SC state, due to redistribution of the $f$-electron charge within the unit cell \cite{Flint:2010hc,Flint:2011cl}. This effect has been well-documented in \PuCoIn\ \cite{Koutroulakis:2014} and also observed in \PuRhIn \cite{RhIn5}, but we were not able to detect a similar appreciable shift of \nuq\ below \Tc\ in \PuCoGa . 

Another pivotal consequence of the composite pair formation is the enhancement of the normal-state \iTone\ near to the SC transition, as the local moments correlate between sites approaching the composite pairing. This results in the interference of the Kondo effect in the two screening channels, giving rise to a relaxation term with a predicted form for the related upturn \mbox{$\left(T_1T\right)^{-1}\propto \left[\ln^2\left(T/T_c\right)+\alpha^2\right]^{-1}$}, where $\alpha$ is a parameter of order $\pi$. \cite{Flint:2008fk} In Fig. \ref{fig:invT1T}, the measured $\left(T_1T\right)^{-1}$ is compared to the predicted term added to the $T$-independent Korringa background at higher temperature (solid red curve). This form certainly captures qualitatively the observed increase in the relaxation nearing \Tc, and it also produces an excellent quantitative agreement with the measured rate, albeit for the relatively low value of $\alpha=0.76$.

\section{Conclusion}
	
Our measurements look to verify and expand upon previous NQR and NMR studies on \PuCoGa\ \cite{Curro:2005fw,Curro:2006,Baek:2010ca}, in an effort to help resolve the puzzle of the role of charge (valence) and spin fluctuations on stabilizing superconductivity in the Pu-115s, as well as in heavy-fermion compounds in general. Specifically, we investigated the NQR properties of the two Ga sites of both NQR active isotopes for a wide temperature range, $T\sim$1.6 K--300 K. The quadrupole frequency behaves anomalously with lowering temperature below $T\simeq 50$ K,  which is attributed to an unusual variation of the EFG. However, both the temperature evolution of the relaxation rate and its ratio for different isotopes fail to produce the signatures expected in the case of critical valence fluctuations. Instead, our relaxation rate results corroborate the emergence of strong in-plane spin fluctuations close to \Tc , which are believed to be key for the the SC pairing, and are found to be similar to several other unconventional superconductors. The striking similarity between the relaxation temperature dependence in \PuCoGa\ and \PuCoIn , especially, suggests a common nature for the fluctuations in these materials. Nevertheless, the presence and effect of normal-state charge fluctuations can not be excluded. Considering the findings of ultrasound spectroscopy in conjunction with our NQR results, it is possible that fluctuations of the 5$f$ valence due to hybridization with the conduction electrons carry strong spin fluctuations of similar nature. Then, the sensitivity of the different measurements to any of the two species would depend on the relevant time- and energy-scales. Further studies are necessary to investigate and clarify the possible connection.

Lastly, the key prediction of the composite pairing theory for a sharp NQR frequency shift upon entering the SC state \cite{Flint:2010hc,Flint:2011cl} could not be verified. 

\section{Acknowledgements}

This work was performed under the auspices of the U.S. Department of Energy, Office of Basic Energy Sciences, Division of Materials Sciences and Engineering. P.H.T. and J.N.M were supported through the Los Alamos Laboratory Directed Research and Development program. G.K. and H.Y. acknowledge support from the Glenn T. Seaborg Institute.
\\

\bibliographystyle{apsrev4-1}

\end{document}